\documentclass{article}
\usepackage{latexsym}
\begin{document}
\begin{large}

\begin{center}
\textbf{\LARGE On the Solvability of Magnetic Differential Equations}
\end{center}

\begin{center}
W. Engelhardt\footnote{Fasaneriestrasse 8, D-80636 M\"{u}nchen, Germany, e-mail: wolfgangw.engelhardt@t-online.de}, retired from: 
\end{center} 

\begin{center}
Max-Planck-Institut f\"{u}r Plasmaphysik, IPP-Euratom Association, 
Garching, Germany
\end{center}
 
\vspace{1. cm} 

\noindent \textbf{Abstract}

\noindent The calculation of both resistive and ideal plasma equilibria amounts to 
solving a number of magnetic differential equations which are of the type 
$\vec {B}\cdot \nabla \Phi =s$. We apply the necessary and sufficient 
criterion for the existence of the potential $\Phi$ and find that a static 
equilibrium configuration of a magnetically confined plasma does not exist 
in axi-symmetric toroidal geometry. 

\vspace{1. cm}

\noindent \textbf {Keywords:} 
\begin{itemize}
\item Magnetohydrodynamics 
\item Magnetic plasma confinement 
\item First-order partial differential equations
\end{itemize}

\vspace{1. cm}

\noindent \textbf{I Introduction }

\noindent The term `magnetic differential equation' was coined by Kruskal and Kulsrud 
[1] in 1958. It arose from an attempt to model a `resistive plasma 
equilibrium' with the equations:
\begin{equation}
\label{eq1}
\vec {j}\times \vec {B}=\nabla p
\end{equation}
\begin{equation}
\label{eq2}
rot\,\vec {B}=\mu _0 \vec {j}
\end{equation}
\begin{equation}
\label{eq3}
div\,\vec {B}=0
\end{equation}
\begin{equation}
\label{eq4}
\vec {E}+\vec {v}\times \vec {B}=\eta \,\vec {j}
\end{equation}
\begin{equation}
\label{eq5}
rot\,\vec {E}=0
\end{equation}
From (\ref{eq5}) follows $\vec {E}=-\nabla \Phi $ and from (\ref{eq4}) after taking the 
scalar product with $\vec {B}$:
\begin{equation}
\label{eq6}
\vec {B}\cdot \nabla \Phi =-\eta \,\vec {j}\cdot \vec {B}=s
\end{equation}
This equation was called a magnetic differential equation. Newcomb [2] 
formulated a criterion for its solvability in 1959:
\begin{equation}
\label{eq7}
\oint {\frac{s\,dl}{\left| {\vec {B}} \right|}=0} 
\end{equation}
where the integration has to be carried out around closed field lines.

Equations (1 - 5) are the basis for calculating the classical diffusive 
particle losses in a magnetically confined plasma. Pfirsch and Schl\"{u}ter 
[3] used a magnetic model field which did not exactly satisfy equations (1 - 
2), but they obtained, nevertheless, an estimate for the diffusion 
coefficient in a toroidal configuration. Later on Maschke [4] generalized 
their results for a true equilibrium configuration and confirmed the order 
of magnitude of the predicted losses. The present state of the art to 
calculate collisional transport in tokamaks including the effect of trapped 
particles can be found, e.g., in [5].

In this paper we analyze again the solvability of (\ref{eq6}) and find that 
Newcomb's criterion (\ref{eq7}) is necessary, but not sufficient. In general, the 
integral $\oint {\nabla \phi \cdot d\vec {l}} $ taken around any closed 
loop, which does not surround the central hole of the torus, must vanish, if 
the potential $\phi $ is to exist. If we restrict ourselves to axi-symmetric 
configurations (e.g. tokamak), we see that (\ref{eq7}) is only applied to a special 
class of closed loops, namely to the nested contours of the magnetic 
surfaces. In Section IV we apply the necessary \underline {and} sufficient 
condition (\ref{eq5}) on (\ref{eq6}) and find that the inhomogeneous part $s$ of the 
magnetic differential equation (\ref{eq6}) must vanish. Since $s$ is given by other 
requirements, the condition for the existence of the potential cannot be 
satisfied.

In Section V we reduce the force balance (\ref{eq1}) to a magnetic differential 
equation which is of the same structure as (\ref{eq6}). It turns out that the 
criterion for its solvability can also not be satisfied in general. The 
reason is that the transformation properties of the cross-product in (\ref{eq1}) are 
at variance with the transformation properties of the pressure gradient. As 
a consequence we come to the conclusion that the assumed stationary state of 
a magnetically confined plasma as described by (1 - 5) does not exist.

\vspace{0.6 cm} 

\noindent \textbf{II Equilibrium configuration }

\noindent We introduce a cylindrical coordinate system with unit vectors $\left( {\vec 
{e}_R ,\,\vec {e}_\varphi ,\,\vec {e}_Z } \right)$ and take the toroidal 
angle as an ignorable coordinate (axi-symmetry). The magnetic field 
satisfying (\ref{eq3}) may be written as: 
\begin{equation}
\label{eq8}
\vec {B}=\vec {B}_p +\vec {B}_\varphi =\nabla \psi \times \nabla \varphi 
+\frac{F}{R}\,\vec {e}_\varphi 
\end{equation}
where $\psi $ is the poloidal magnetic flux function. From (\ref{eq1}) and the 
poloidal component of (\ref{eq2}) follows then: 
\begin{equation}
\label{eq9}
p=p\left( \psi \right)\;,\quad F=F\left( \psi \right)
\end{equation}
and the current density may be expressed by
\begin{equation}
\label{eq10}
\vec {j}=\frac{F'}{\mu _0 }\vec {B}_p +\left( {R{\kern 1pt}p'+\frac{FF'}{\mu 
_0 \, R }} \right)\vec {e}_\varphi 
\end{equation}
where $'$ denotes differentiation with respect to $\psi $ . Inserting this 
into the toroidal component of (\ref{eq2}) yields the 
L\"{u}st-Schl\"{u}ter-Grad-Rubin-Shafranov equation [6]:
\begin{equation}
\label{eq11}
\frac{\partial ^2\psi }{\partial R^2}-\frac{1}{R}\frac{\partial \psi 
}{\partial R}+\frac{\partial ^2\psi }{\partial Z^2}=\Delta ^\ast \psi 
=-\left( {R^2\mu _0 p'+FF'} \right)
\end{equation}
This equation is usually solved numerically with suitable boundary 
conditions to find the equilibrium magnetic field in axi-symmetric plasma 
configurations. There exist also analytical solutions (see Appendix A) which 
contain free parameters so that the solution can be adapted to realistic 
boundary conditions. Recently equation (\ref{eq11}) has been modified [7] to include 
approximately the effect of magnetic islands which break axi-symmetry. At 
this point, however, we assume that a solution of (\ref{eq11}) is given for a 
particular choice of $p'$ and $FF'$.

\vspace{0.6 cm}
 

\noindent \textbf{III Standard method for calculating the electric field and the 
velocity field}

\noindent The method used in [3] and [4] for calculating the electric field is 
straightforward. Because of (\ref{eq5}) the electric field may be written as:
\begin{equation}
\label{eq12}
\vec {E}=\vec {E}_p +\vec {E}_\varphi =-\nabla \phi +\frac{U}{2\pi R}\,\vec 
{e}_\varphi 
\end{equation}
where $U$ is the loop voltage produced by a transformer. Substituting this 
into (\ref{eq4}) and taking the scalar product with $\vec {B}$ yields the magnetic 
differential equation: 
\begin{equation}
\label{eq13}
-\vec {B}_p \cdot \nabla \phi =\left( {\eta {\kern 1pt}j_\varphi -E_\varphi 
} \right)\frac{F}{R}+\frac{\eta \,F'}{\mu _0 }B_p^2 
\end{equation}
with the solution:
\begin{equation}
\label{eq14}
\phi =\phi _0 \left( \psi \right)-\int\limits_{l_0 }^l {\left[ {\left( {\eta 
{\kern 1pt}j_\varphi -E_\varphi } \right)\frac{F}{R}+\frac{\eta \,F'}{\mu _0 
}B_p^2 } \right] \frac{R\,dl}{\left| {\nabla \psi } \right|}} 
\end{equation}
where $dl$ denotes a line element on a contour of a magnetic surface $\psi 
=const$. In order to obtain a single-valued potential, the closed integral 
around a magnetic surface must vanish:
\begin{equation}
\label{eq15}
\oint {\left[ {\left( {\eta {\kern 1pt}j_\varphi -E_\varphi } 
\right)\frac{F}{R}+\frac{\eta \,F'}{\mu _0 }B_p^2 } \right]} 
\frac{R\,dl}{\left| {\nabla \psi } \right|}=0
\end{equation}
This equation reflects Newcomb's condition (\ref{eq7}) and puts a constraint on the 
choice of the functions $p'$ and $F'$ to be used in (\ref{eq11}).

The velocity field is obtained by taking the cross-product of (\ref{eq4}) with $\vec 
{B}$: 
\begin{equation}
\label{eq16}
\vec {v}=\alpha \left( {R,\,Z} \right)\vec {B}+{\left( {\vec {E}-\eta \,\vec 
{j}} \right)\times \vec {B}} \mathord{\left/ {\vphantom {{\left( {\vec 
{E}-\eta \,\vec {j}} \right)\times \vec {B}} {B^2}}} \right. 
\kern-\nulldelimiterspace} {B^2}
\end{equation}
The arbitrary function $\alpha $ may be expressed by the divergence of the 
velocity, thus resulting in a second magnetic differential equation because 
of (\ref{eq3}):
\begin{equation}
\label{eq17}
\vec {B}_p \cdot \nabla \alpha =div\left( {\vec {v}-{\left( {\vec {E}-\eta 
\,\vec {j}} \right)\times \vec {B}} \mathord{\left/ {\vphantom {{\left( 
{\vec {E}-\eta \,\vec {j}} \right)\times \vec {B}} {B^2}}} \right. 
\kern-\nulldelimiterspace} {B^2}} \right)
\end{equation}
The solution is:
\begin{equation}
\label{eq18}
\alpha =\alpha _0 \left( \psi \right)+\int\limits_{l_0 }^l {div\left( {\vec 
{v}-{\left( {\vec {E}-\eta \,\vec {j}} \right)\times \vec {B}} 
\mathord{\left/ {\vphantom {{\left( {\vec {E}-\eta \,\vec {j}} \right)\times 
\vec {B}} {B^2}}} \right. \kern-\nulldelimiterspace} {B^2}} \right)} 
\frac{R\,dl}{\left| {\nabla \psi } \right|}
\end{equation}
with the integrability condition:
\begin{equation}
\label{eq19}
\oint {div\,\vec {v}} \frac{R\,dl}{\left| {\nabla \psi } \right|}=\oint 
{div\left( {{\left( {\vec {E}-\eta \,\vec {j}} \right)\times \vec 
{B}} \mathord{\left/ {\vphantom {{\left( {\vec {E}-\eta \,\vec {j}} 
\right)\times \vec {B}} {B^2}}} \right. \kern-\nulldelimiterspace} {B^2}} 
\right)} \frac{R\,dl}{\left| {\nabla \psi } \right|}
\end{equation}
which is usually expressed in the form:
\begin{equation}
\label{eq20}
\int {\oint {div\,\vec {v}} } \;\frac{R\,dl}{\left| {\nabla \psi } 
\right|}\;d\psi =\oint {\left( {\vec {E}\times \vec {B}-\eta \,\nabla p} 
\right)\cdot \nabla \psi } \;\frac{R\,dl}{B^2\left| {\nabla \psi } \right|}
\end{equation}
by application of Gauss' theorem.

It appears now that the problem is solved, first by prescribing a flux 
function satisfying (\ref{eq11}) and (\ref{eq15}), a divergence of the velocity subject to 
the constraint (\ref{eq20}), and secondly, by evaluating (\ref{eq14}) and (\ref{eq18}). The 
solutions may be substituted into (\ref{eq12}) and (\ref{eq16}) to yield the vector fields 
$\vec {E}$ and $\vec {v}$. The integration functions $\phi _0 $ and $\alpha 
_0 $ describe an arbitrary rotation in poloidal and toroidal direction. In 
the following it is shown that this expectation is not justified, since (\ref{eq15}) 
and (\ref{eq20}) turn out to be necessary, but not sufficient conditions. 

\vspace{0.6 cm} 

\noindent \textbf{IV Application of the necessary and sufficient criterion for the 
existence of the potential}

\noindent Before we apply the necessary and sufficient condition (\ref{eq5}) for the existence 
of the potential $\phi $, we write (\ref{eq13}) in the form:
\begin{equation}
\label{eq26}
E_R \,B_R +E_Z \,B_Z =s \quad , \quad \quad s=\left( {\eta {\kern 1pt}j_\varphi -E_\varphi} \right)\frac{F}{R}+\frac{\eta \,F'}{\mu _0 }B_p^2 
\end{equation}
and take the gradient of this equation:
\begin{eqnarray}
\label{eq27}
B_R \frac{\partial E_R }{\partial R}+B_Z \frac{\partial E_Z }{\partial 
R}+E_R \frac{\partial B_R }{\partial R}+E_Z \frac{\partial B_Z }{\partial 
R}=\frac{\partial s}{\partial R} \nonumber \\
B_R \frac{\partial E_R }{\partial Z}+B_Z \frac{\partial E_Z }{\partial 
Z}+E_R \frac{\partial B_R }{\partial Z}+E_Z \frac{\partial B_Z }{\partial 
Z}=\frac{\partial s}{\partial Z}
\end{eqnarray}
Together with (\ref{eq26}), (\ref{eq3}), and the necessary and sufficient condition (\ref{eq5}):
\begin{equation}
\label{eq28}
\frac{\partial E_R }{\partial Z}=\frac{\partial E_Z }{\partial R}
\end{equation}
one obtains from (\ref{eq27}) two magnetic differential equations for the electric 
field components:
\begin{equation}
\label{eq29}
\vec {B}_p \cdot \nabla \left( {\frac{E_R }{R\;B_Z }} 
\right)=\frac{1}{R}\frac{\partial }{\partial R}\left( {\frac{s}{B_Z }} 
\right)\;,\quad \vec {B}_p \cdot \nabla \left( {\frac{E_Z }{R\;B_R }} 
\right)=\frac{1}{R}\frac{\partial }{\partial Z}\left( {\frac{s}{B_R }} 
\right)
\end{equation}
In analogy to (\ref{eq14}) they have the solutions:
\begin{eqnarray}
\label{eq30}
E_R =R\,B_Z \left[ {\int\limits_{l_0 }^l {\frac{\partial }{\partial R}\left( 
{\frac{s}{B_Z }} \right) \frac{dl}{\left| {\nabla \psi } \right|} 
 +f_1 \left( \psi \right)} } \,
\right]\; \nonumber \\
E_Z =R\,B_R \left[ {\int\limits_{l_0 }^l {\frac{\partial }{\partial Z}\left( 
{\frac{s}{B_R }} \right) \frac{dl}{\left| {\nabla \psi } \right|} 
+f_2 \left( \psi \right)} }\, 
\right]
\end{eqnarray}
Inserting this into (\ref{eq26}) yields:
\begin{equation}
\label{eq31}
f_1 \left( \psi \right)+f_2 \left( \psi \right)+\int\limits_{l_0 }^l {\left[ 
{\frac{\partial }{\partial R}\left( {\frac{s}{B_Z }} \right)+\frac{\partial 
}{\partial Z}\left( {\frac{s}{B_R }} \right)} \right] \frac{dl}{\left| {\nabla \psi } \right|} =\frac{s}{R\,B_R B_Z }} 
\end{equation}
Since the lower bound $l_0 $ of the integral may be chosen arbitrarily close 
to the upper bound $l$, the integral can be made to vanish so that equation 
(\ref{eq31}) leads to the condition:
\begin{equation}
\label{eq32}
f_2 \left( \psi \right)=-f_1 \left( \psi \right)=\phi _0 '\,\left( \psi 
\right)\quad ,\quad \quad s=0
\end{equation}
as the right-hand-side of (\ref{eq31}) is not a function of $\psi$ alone. Result (\ref{eq32}) can also be deduced from the transformation properties of the electric field vector whose components are given in (\ref{eq30}). This is demonstrated in Appendix B. 

The local condition (\ref{eq32}) is more restrictive than Newcomb´s integral condition (\ref{eq15}), and it is not compatible with $s$ as given in (\ref{eq26}). Hence, we must conclude that the potential does not exist.

\vspace{0.6 cm} 

\noindent \textbf{V Magnetic differential equations and the force balance}

\noindent In case of the static force balance (\ref{eq1}) scalar multiplication with the 
magnetic field yields the homogeneous magnetic differential equation:
\begin{equation}
\label{eq38}
\vec {B}\cdot \nabla p=0
\end{equation}
which satisfies condition (\ref{eq32}) so that an ideal equilibrium with $\eta =0$ 
should be possible. The necessary and sufficient condition for the existence 
of the pressure `potential' is, however, that the curl of (\ref{eq1}) vanishes: 
$\left( {\vec {B}\cdot \nabla } \right)\vec {j}=\left( {\vec {j}\cdot \nabla 
} \right)\vec {B}$, which again leads to an inhomogeneous magnetic 
differential equation for the toroidal component of the current density 
together with (\ref{eq2}): 
\begin{equation}
\label{eq39}
\vec {B}_p \cdot \nabla \left( {\frac{j_\varphi }{R}} \right)=-\frac{2{\kern 
1pt}B_\varphi j_R }{R^2}
\end{equation}
As shown in the previous Section, it has only the solution $j_\varphi 
=R\,f\left( \psi \right)$ so that the radial component of the current 
density must vanish.

This may also be shown by considering the so called `force-free' situation 
where the pressure gradient vanishes and (\ref{eq11}) becomes:
\begin{equation}
\label{eq40}
\Delta ^\ast \psi =-FF'=-g\left( \psi \right)
\end{equation}
Applying Stokes's theorem on this equation by integrating the toroidal current density over the area enclosed by a magnetic surface one has:
\begin{equation}
\label{eq41}
\oint {\frac{\left| {\nabla \psi } 
\right|}{R}\;dl=\mathop{{\int\!\!\!\int}\mkern-21mu \bigcirc} {{\kern 
1pt}g\left( \psi \right)} } \;\frac{dR\;dZ}{R}
\end{equation}
With:
\begin{equation}
\label{eq42}
\psi =\int {\frac{F}{g\left( F \right)}\;dF\quad ,\quad \quad \nabla \psi 
=\frac{F}{g}\;\nabla F} 
\end{equation}
one obtains from (\ref{eq40}):
\begin{equation}
\label{eq43}
\Delta ^\ast F=-\frac{g^2}{F}-\frac{g}{F}\,\left| {\nabla F} 
\right|^2\frac{d}{dF}\left( {\frac{F}{g}} \right)
\end{equation}
Stoke's theorem applied on this equation gives:
\begin{equation}
\label{eq44}
\oint {\frac{\left| {\nabla F} \right|}{R}\;dl} 
=\mathop{{\int\!\!\!\int}\mkern-21mu \bigcirc} {\left( 
{\frac{g^2}{F}+\frac{g}{F}\,\left| {\nabla F} \right|^2\frac{d}{dF}\left( 
{\frac{F}{g}} \right)} \right)} \;\frac{dR\;dZ}{R}{\kern 1pt}
\end{equation}
Substitution of (\ref{eq42}) into (\ref{eq41}) yields on the other hand:
\begin{equation}
\label{eq45}
\frac{F}{g}\oint {\frac{\left| {\nabla F} \right|}{R}\;dl} 
=\mathop{{\int\!\!\!\int}\mkern-21mu \bigcirc} {{\kern 
1pt}g\;\frac{dR\;dZ}{R}} 
\end{equation}
Elimination of the line integral over the poloidal current density on the 
left-hand sides of (\ref{eq44}) and (\ref{eq45}), and using (\ref{eq42}) again results in an integral equation:
\begin{equation}
\label{eq46}
g\;\mathop{{\int\!\!\!\int}\mkern-21mu \bigcirc} {g\;\frac{dR\;dZ}{R}} 
=F\;\mathop{{\int\!\!\!\int}\mkern-21mu \bigcirc} {{\kern 1pt}\left( 
{\frac{g^2}{F}+\frac{1}{F^3}\,\left( {g^2-F\;\frac{dg}{d\psi }} 
\right)\left| {\nabla \psi } \right|^2} \right)} \;\frac{dR\;dZ}{R}
\end{equation}
which can only be satisfied for $g=FF'=0$ in agreement with (\ref{eq32}) and (\ref{eq39}). We demonstrate this explicitly in Appendix A by adopting a `Soloviev solution' [8] with $g=const$.

In view of this result it becomes doubtful whether the condition $p=p\left( 
\psi \right)$ resulting from (\ref{eq38}) in axi-symmetry defines a scalar pressure, or, in other words, whether the cross-product 
$\vec {j}\times \vec {B}$, which is in principle an antisymmetric second-rank tensor, can have the same 
transformation properties as the polar vector field $\nabla p$, at least 
under certain circumstances.

In order to investigate this question we apply the Laplace operator in 
Cartesian coordinates on (\ref{eq38}):
\begin{eqnarray}
\vec {B}\cdot \nabla \left( {\Delta p} \right)+\nabla p\cdot \Delta \vec 
{B} \quad \quad \quad\quad \quad \quad\quad \quad \quad \quad \quad \quad\quad \quad \quad\quad \quad \quad\quad \quad \quad\quad \quad \quad \nonumber \\
\label{eq47} +2\left( {\frac{\partial B_x }{\partial x}\frac{\partial 
^2p}{\partial x^2}+\frac{\partial B_y }{\partial y}\frac{\partial 
^2p}{\partial y^2}+\frac{\partial B_z }{\partial z}\frac{\partial 
^2p}{\partial z^2}+k_1 \frac{\partial ^2p}{\partial x\partial y}+k_2 
\frac{\partial ^2p}{\partial x\partial z}+k_3 \frac{\partial ^2p}{\partial 
y\partial z}} \right)=0 \\  
k_1 =\frac{\partial B_x }{\partial 
y}+\frac{\partial B_y }{\partial x}\;,\quad k_2 =\frac{\partial B_x 
}{\partial z}+\frac{\partial B_z }{\partial x}\;,\quad k_3 =\frac{\partial 
B_y }{\partial z}+\frac{\partial B_z }{\partial y}
\quad \quad \quad\quad \quad \nonumber 
\end{eqnarray}
The gradient of (\ref{eq38}):
\begin{eqnarray}
\frac{\partial }{\partial x}\left( {B_x \frac{\partial p}{\partial x}+B_y 
\frac{\partial p}{\partial y}+B_z \frac{\partial p}{\partial z}} \right)=0 
\nonumber \\ 
\label{eq48}
\frac{\partial }{\partial y}\left( {B_x \frac{\partial p}{\partial x}+B_y 
\frac{\partial p}{\partial y}+B_z \frac{\partial p}{\partial z}} \right)=0 \\ 
 \nonumber \frac{\partial }{\partial z}\left( {B_x \frac{\partial p}{\partial x}+B_y 
\frac{\partial p}{\partial y}+B_z \frac{\partial p}{\partial z}} \right)=0 
\end{eqnarray}
yields three equations which can be used to eliminate the mixed derivatives 
in (\ref{eq47}). The bracket may then be written in the form:
\begin{eqnarray} 
\left( {2\frac{\partial B_x }{\partial x}+l_x B_x } \right)\frac{\partial 
^2p}{\partial x^2}+\left( {2\frac{\partial B_y }{\partial y}+l_y B_y } 
\right)\frac{\partial ^2p}{\partial y^2}+\left( {2\frac{\partial B_z 
}{\partial z}+l_z B_z } \right)\frac{\partial ^2p}{\partial z^2}+\nabla 
p\cdot \left( {\vec {l}\cdot\nabla } \right)\vec {B} \nonumber \\ \label{eq49} \\
 \vec {l}=\left( {\frac{k_3 B_x }{B_y B_z }-\frac{k_1 }{B_y 
}-\frac{k_2 }{B_z }} \right)\;\vec {e}_x +\left( {\frac{k_2 B_y }{B_x B_z 
}-\frac{k_1 }{B_x }-\frac{k_3 }{B_z }} \right)\;\vec {e}_y +\left( 
{\frac{k_1 B_z }{B_x B_y }-\frac{k_2 }{B_x }-\frac{k_3 }{B_y }} 
\right)\;\vec {e}_z \nonumber 
\end{eqnarray}
It turns out that this expression does not have the transformation 
properties of a scalar, in contrast to the first two terms in (\ref{eq47}). By 
`scalar' we refer to a quantity which does not change its value, when it is 
expressed in different coordinate systems as a function of space: $p\left( 
{\vec {x}} \right)=p\left( {\vec {x}\,'} \right),\;\vec {x}\,'=\vec {x},\;x_i 
'=a_{ik} {\kern 1pt}x_k $. The reason for the `non-scalar' property of (\ref{eq49}) is that the directed quantity $\vec {l}$ cannot be considered as a vector field which maintains its modulus as a scalar, when it is transformed into a rotated coordinate system. The inner product of the last term in (\ref{eq49}) will, therefore, depend on the orientation of the coordinate system which would not be the case for the invariant inner product $\nabla p\cdot \left( {\vec {B}\cdot \nabla } \right)\vec {B}$, e.g., or for the first two terms in (\ref{eq47}). Similar remarks apply to the first three terms in (\ref{eq49}) which resemble 
the Laplacian of the pressure, but the second derivatives have coefficients which are all different so that the sum of these terms is not invariant, as compared to the Laplacian of a scalar field. Consequently, equation (\ref{eq49}) leads to an incongruity, when it is transformed into a rotated coordinate system, as it will depend explicitly on the rotational angle.

In order to show this we choose a coordinate system which is rotated around 
the $y$-axis by an angle $\alpha $:
\begin{equation}
\label{eq50}
x=x'\cos \alpha -z'\sin \alpha \;,\quad z=x'\sin \alpha +z'\cos \alpha 
\;,\quad y=y'
\end{equation}
The transformation rules are:
\begin{equation}
\label{eq51}
\frac{\partial }{\partial x}=\cos \alpha \,{\kern 1pt}\frac{\partial 
}{\partial x'}-\sin \alpha \frac{\partial }{\partial z'}\;,\quad 
\frac{\partial }{\partial z}=\sin \alpha \frac{\partial }{\partial x'}+\cos 
\alpha \frac{\partial }{\partial z'}\;,\quad \frac{\partial }{\partial 
y}=\frac{\partial }{\partial y'}
\end{equation}
Applying these to (\ref{eq49}) one finds that the transformed expression contains 
not only the components of the magnetic field and of the pressure gradient 
in the primed system, but in addition the rotational angle $\alpha $. This 
may be most readily verified by transforming the coefficient $l_y $. It 
should be invariant, since the $y$-component of the pressure gradient does not 
change under the assumed rotation so that the second term in (\ref{eq49}) should be the same in the rotated system. Instead one obtains:
\begin{equation}
\label{eq52}
l_y '=\frac{\left( {k_2 'B_{y'} -k_1 'B_{z'} -k_3 'B_{x'} } \right)\cos 
2\alpha +\left( {k_1 'B_{x'} -k_3 'B_{z'} +\left( {\frac{\partial B_{z'} 
}{\partial z'}- {\frac{\partial B_{x'} }{\partial x'}} } 
\right)B_{y'} } \right)\sin 2\alpha }{\left( {B_{x'} \cos \alpha -B_{z'} 
\sin \alpha } \right)\;\left( {B_{x'} \sin \alpha +B_{z'} \cos \alpha } 
\right)}
\end{equation}
where the $k_n '$ are defined as in (\ref{eq47}) in terms of primed derivatives of the primed field components. For $\alpha =0$ one returns to the expression 
$l_y $ as given in (\ref{eq49}). Evidently, the expression (\ref{eq49}) does not transform like a scalar field which is only the case when the pressure itself, as defined by (\ref{eq38}) in the form $p\left( \psi \right)$, is not a scalar, 
contrary to our assumption. In Appendix A we show this explicitly for the 
pressure as given by a Soloviev solution.

From the transformation properties of (\ref{eq49}), which are a consequence of the hypothetical equation (\ref{eq1}), we infer that a static equilibrium configuration does not exist, since the transformation properties of the vector 
cross-product are incompatible with those of the gradient of a scalar 
pressure.

\vspace{0.6 cm} 

\noindent \textbf{Conclusion}

\noindent The stationary equilibrium equations (1 - 5) of a magnetically confined 
plasma may be formulated in terms of magnetic differential equations. We 
have shown that these equations have ambiguous solutions in axi-symmetry, 
unless their inhomogeneous part vanishes. Furthermore, due to the 
transformation properties of the cross-product $\vec {j}\times \vec {B}$, it 
cannot be set equal to the gradient of a scalar pressure. As a consequence, 
the set of equations (1 - 5) is not solvable. 

The time dependent terms which are omitted in the stationary 
magneto-hydro-dynamic model can apparently not be neglected. Consequently, 
magnetic confinement of a plasma in toroidal geometry leads inevitably to 
temporal changes of the pressure and the electromagnetic field. This may not 
be necessarily a unidirectional temporal evolution, but turbulent 
fluctuations could lead to an average `quasi-stationary' state, which would 
not be strictly axi-symmetric any longer.

\vspace{0.6 cm} 

\noindent \textbf{Acknowledgments}

\noindent The author is grateful to Professor Pfirsch who carefully read the 
manuscript and spotted an error in Section IV which subsequently was 
corrected. Professor Schl\"{u}ter pointed out in discussions that -- 
according to his own analysis of the Vlasov equation -- the static force 
balance (\ref{eq1}) cannot be satisfied when an isotropic pressure is assumed.

\vspace{1. cm} 

\noindent \textbf{Appendix A}
\setcounter{equation}{0} 
\renewcommand{\theequation}{A.\arabic{equation}} 

\noindent Soloviev has constructed a flux function [8]:
\begin{equation}
\label{eq53}
\psi =\frac{1}{2}\left( {c_0 {\kern 1pt}R^2+b\,R_0^2 } \right) Z^2+\frac{a-c_0 
}{8}\left( {R^2-R_0^2 } \right)^2
\end{equation}
which substituted into (\ref{eq8}) yields the poloidal magnetic field components:
\begin{equation}
\label{eq54}
B_R =-\frac{Z}{R}\left( {c_0 {\kern 1pt}R^2+b\,R_0^2 } \right)\;,\quad B_Z 
=c_0 {\kern 1pt}Z^2+\frac{a-c_0 }{2}\left( {R^2-R_0^2 } \right)
\end{equation}
When these are inserted into the toroidal component of (\ref{eq2}) one obtains:
\begin{equation}
\label{eq55}
\mu _0 j_\varphi =-a\,R-\frac{b\,R_0^2 }{R}
\end{equation}
so that (\ref{eq10}) and (\ref{eq11}) are satisfied with:
\begin{equation}
\label{eq56}
p'=-\mu _0 {\kern 1pt}a\;,\quad FF'=-b\,R_0^2 
\end{equation}
Recently, expressions similar to (\ref{eq53}) were published in [9], which have more 
adjustable parameters. They allow to model tokamak plasma configurations 
more realistically than (\ref{eq53}).

With $g=-b{\kern 1pt}R_0^2 $ equation (\ref{eq46}) becomes: 
\begin{equation}
\label{eq57}
g^2\mathop{{\int\!\!\!\int}\mkern-21mu \bigcirc} {\frac{dR\;dZ}{R}} 
=g^2{\kern 1pt}F\,\mathop{{\int\!\!\!\int}\mkern-21mu \bigcirc} {\left( 
{\frac{1}{F}+\frac{\left| {\nabla \psi } \right|^2}{F^3}} 
\right)\frac{dR\;dZ}{R}} 
\end{equation}
and from (\ref{eq56}) follows: 
\begin{equation}
\label{eq58}
F=\sqrt {F_0^2 -2{\kern 1pt}b{\kern 1pt}R_o^2 {\kern 1pt}\psi } 
\end{equation}
Converting the left-hand-side of (\ref{eq57}) into a line integral and performing a 
partial integration on the first term of the right-hand-side of (\ref{eq57}) yields 
with (\ref{eq58}):
\begin{equation}
\label{eq59}
g^2\int\limits_{R_1 }^{R_2 } {\frac{Z\left( {R,\psi } \right)}{R}\,dR} 
=g^2{\kern 1pt}F\left[ {\int\limits_{R_1 }^{R_2 } {\frac{Z\left( {R,\psi } 
\right)}{F\,R}\,dR+\mathop{{\int\!\!\!\int}\mkern-21mu \bigcirc} {\left( 
{-\frac{b{\kern 1pt}R_0^2 {\kern 1pt}Z}{F^3}{\kern 1pt}\frac{\partial \psi 
}{\partial Z}+\frac{\left| {\nabla \psi } \right|^2}{F^3}} 
\right)\frac{dR\,dZ}{R}} } } \right]
\end{equation}
where $R_1 $ and $R_2 $ are the points where a magnetic surface cuts the 
mid-plane $Z=0$. The $Z$ - coordinate on a magnetic surface is expressed as 
a function of $R$ and $\psi $ with (\ref{eq53}). Collecting terms equation (\ref{eq59}) 
becomes with (\ref{eq8}):
\begin{equation}
\label{eq60}
0=F{\kern 1pt}g^2\mathop{{\int\!\!\!\int}\mkern-21mu \bigcirc} {\left( 
{b{\kern 1pt}R_0^2 {\kern 1pt}Z{\kern 1pt}B_R +R\left( {B_R^2 +B_Z^2 } 
\right)} \right)\,\frac{dR\,dZ}{F^3}} 
\end{equation}
Inserting the magnetic field components as given in (\ref{eq54}) with $a=0$ one 
finds that the double integral over the cross-section of the plasma inside a 
magnetic surface does not vanish which requires then $g=0$ to satisfy (\ref{eq60}). 
This result was already expected from the condition (\ref{eq32}) and the magnetic 
differential equation (\ref{eq39}).

Soloviev's solution may help to understand the conclusion following from 
(\ref{eq52}) that $\vec {B}\cdot \nabla p=0$ does not define a scalar pressure. When 
the first expression in (\ref{eq56}) is integrated, one obtains for the pressure:
\begin{equation}
\label{eq61}
p=p_0 -\mu _0 {\kern 1pt}a\,\psi 
\end{equation}
Because of $\vec {B}=rot\,\vec {A}$ the flux function $\psi $ in (\ref{eq8}) is 
related to the toroidal component of the vector potential: $\psi =R{\kern 
1pt}A_\varphi $. This expression may be considered as the $Z$ - component of 
the vector field $R{\kern 1pt}{\kern 1pt}\vec {e}_R \times A_\varphi {\kern 
1pt}\vec {e}_\varphi $. An arbitrary function of a single vector component 
does, however, not transform like a scalar field. Writing equation (\ref{eq61}) in 
the form:
\begin{equation}
\label{eq62}
p=p_0 -\mu _0 {\kern 1pt}a{\kern 1pt}R{\kern 1pt}A_\varphi = p_0 -\mu _0 {\kern 
1pt}a\left( {x{\kern 1pt}A_y -y{\kern 1pt}A_x } \right)
\end{equation}
and transforming this expression into a coordinate system which is rotated 
around the $y$ - axis by an angle $\alpha $ as in Section V:
\begin{equation}
\label{eq63}
\begin{array}{l}
 \quad \quad x=x'\cos \alpha -z'\sin \alpha \;,\quad z=x'\sin \alpha +z'\cos 
\alpha \;,\quad y=y' \\ 
 \\ 
 A_x =A_{x'} \cos \alpha -A_{z'} \sin \alpha \;,\quad A_z =A_{x'} \sin 
\alpha +A_{z'} \cos \alpha \;,\quad A_y =A_{y'} \\ 
 \end{array}
\end{equation}
one obtains:
\begin{equation}
\label{eq64}
p=p_0 -\mu _{0{\kern 1pt}} {\kern 1pt}a\,\left[ {\left( {x'A_{y'} -y'A_{z'} 
} \right)\cos \alpha +\left( {y'A_{z'} -z'A_{y'} } \right)\sin \alpha } 
\right]
\end{equation}
This expression contains not only the coordinates and the components of the 
vector potential in the primed system, but in addition the rotational angle 
$\alpha $. Hence, the pressure obtained from (\ref{eq61}) does not transform like a 
scalar.

\vspace{1. cm} 

\noindent \textbf{Appendix B}
\setcounter{equation}{0} 
\renewcommand{\theequation}{B.\arabic{equation}} 

\noindent The derivation of the condition $s=0$ in Section IV rested on a particular 
choice of the lower bound of the integral (\ref{eq31}). More generally, result (\ref{eq32}) 
follows also from the vector character of the electric field whose 
components are given in (\ref{eq30}) in cylindrical coordinates. If one formulates 
equation (\ref{eq26}) in spherical coordinates ($R=r\sin \theta \,,\;Z=r\cos \theta 
\,,\;\varphi =\varphi )$:
\begin{equation}
\label{eq69}
E_r \,B_r +E_\theta \,B_\theta =s
\end{equation}
the method applied in Section IV yields now expressions for the field components 
in spherical coordinates. Taking the gradient of (\ref{eq69}) one obtains with (\ref{eq3}) 
and (\ref{eq5}) two magnetic differential equations for the electric field 
components:
\begin{equation}
\label{eq70}
\vec {B}_p \cdot \nabla \left( {\frac{E_r }{R\,B_\theta }} 
\right)=\frac{1}{R\,r}\frac{\partial }{\partial r}\left( 
{\frac{s\,r}{B_\theta }} \right)\;,\quad \vec {B}_p \cdot \nabla \left( 
{\frac{E_\theta }{R\,B_r }} \right)=\frac{1}{R\,r}\frac{\partial }{\partial 
\theta }\left( {\frac{s}{B_r }} \right)
\end{equation}
which have the formal solutions:
\begin{eqnarray}
\label{eq71}
E_r =R\,B_\theta \,\left[ {\int\limits_{l_0 }^l {\frac{1}{r}\frac{\partial 
}{\partial r}\left( {\frac{s\,r}{B_\theta }} \right)\frac{dl}{\left| {\nabla 
\psi } \right|}+f_3 \left( \psi \right)} } \right] \nonumber \\
E_\theta =R\,B_r \,\left[ {\int\limits_{l_0 }^l {\frac{1}{r}\frac{\partial 
}{\partial \theta }\left( {\frac{s}{B_r }} \right)\frac{dl}{\left| {\nabla 
\psi } \right|}+f_4 \left( \psi \right)} } \right]
\end{eqnarray}
Writing $E_\theta ={\left( {s-E_r \,B_r } \right)} \mathord{\left/ 
{\vphantom {{\left( {s-E_r \,B_r } \right)} {B_\theta }}} \right. 
\kern-\nulldelimiterspace} {B_\theta }$ and $E_Z ={\left( {s-E_R \,B_R } 
\right)} \mathord{\left/ {\vphantom {{\left( {s-E_R \,B_R } \right)} {B_Z 
}}} \right. \kern-\nulldelimiterspace} {B_Z }$ one obtains for the modulus 
of the electric field from (\ref{eq71}):
\begin{equation}
\label{eq72}
E_r^2 +E_\theta ^2 =R^2B_p^2 \left( {f_3 +I_3 } 
\right)^2+\frac{s^2}{B_\theta ^2 }-\frac{2\,s\,R\,B_r }{B_\theta }\,\left( 
{f_3 +I_3 } \right),\,I_3 =\int\limits_{l_0 }^l {\frac{1}{r}\frac{\partial 
}{\partial r}\left( {\frac{s\,r}{B_\theta }} \right)\frac{dl}{\left| {\nabla 
\psi } \right|}} 
\end{equation}
and from (\ref{eq30}) we have:
\begin{equation}
\label{eq73}
E_R^2 +E_Z^2 =R^2B_p^2 \left( {f_1 +I_1 } \right)^2+\frac{s^2}{B_Z^2 
}-\frac{2\,s\,R\,B_R }{B_Z }\,\left( {f_1 +I_1 } \right),\;I_1 
=\int\limits_{l_0 }^l {\frac{\partial }{\partial R}\left( {\frac{s}{B_Z }} 
\right)\frac{dl}{\left| {\nabla \psi } \right|}} 
\end{equation}
Subtracting both equations one finds:
\begin{eqnarray}
\label{eq74}
\left( {f_1 +I_1 } \right)^2-\left( {f_3 +I_3 } 
\right)^2-\frac{2\,s}{R\,B_p^2 }\left( {\frac{B_R }{B_Z }\,\left( {f_1 +I_1 
} \right)-\frac{B_r }{B_\theta }\,\left( {f_3 +I_3 } \right)} \right) \nonumber \\
\quad \quad \quad \quad +\frac{s^2}{R^2B_p^2 }\left( {\frac{1}{B_Z^2 
}-\frac{1}{B_\theta ^2 }} \right)=0
\end{eqnarray}
If one differentiates this equation along the poloidal magnetic field lines, 
one obtains a linear relationship between the integrals $I_1 $ and $I_3 $:
\begin{eqnarray}
\label{eq75}
k_3 \left( {f_3 +I_3 } \right)=k_1 \left( {f_1 +I_1 } \right)+k_2 \quad \quad \quad \quad \quad \quad \quad \quad \\
k_1 =\vec {B}_p\cdot \nabla I_1 -\vec {B}_p\cdot \nabla \left( {\frac{s\,B_R 
}{R\,B_Z B_p^2 }} \right)\;,\quad \vec {B}_p\cdot \nabla I_1 
=\frac{1}{R}\frac{\partial }{\partial R}\left( {\frac{s}{B_Z }} \right) \nonumber \\
k_3 =\vec {B}_p\cdot \nabla I_3 -\vec {B}_p\cdot \nabla \left( {\frac{s\,B_r 
}{R\,B_\theta B_p^2 }} \right)\;,\quad \vec {B}_p\cdot \nabla I_3 
=\frac{1}{R\,r}\frac{\partial }{\partial r}\left( {\frac{s\,r}{B_\theta }} 
\right) \nonumber \\
k_2 =\vec {B}_p\cdot \nabla \left[ {\frac{s^2}{2\,R^2B_p^2 }\left( 
{\frac{1}{B_Z^2 }-\frac{1}{B_\theta ^2 }} \right)} \right]-\frac{s}{R\,B_p^2 
}\left( {\frac{B_R }{B_Z }\vec {B}_p\cdot \nabla I_1 -\frac{B_r }{B_\theta 
}\vec {B}_p\cdot \nabla I_3 } \right) \nonumber
\end{eqnarray}
Eliminating from (\ref{eq74}) and (\ref{eq75}) the integral $I_3 $ one obtains a quadratic 
equation for $\left( {f_1 +I_1 } \right)$:
\begin{eqnarray}
\label{eq76}
\left( {f_1 +I_1 } \right)^2\left( {1-\frac{k_1^2 }{k_3^2 }} 
\right)-2\,\,\left( {f_1 +I_1 } \right)\,\,\left[ {\frac{k_1 k_2 }{k_3^2 
}+\frac{s}{R\,B_p^2 }\left( {\frac{B_R }{B_Z }-\frac{B_r }{B_\theta 
}\frac{k_1 }{k_3 }} \right)} \right] \nonumber \\
+\frac{2\,s\,B_r }{R\,B_\theta B_p^2 }\frac{k_2 }{k_3 }-\frac{k_2^2 }{k_3^2 
}+\frac{s^2}{R^2B_p^2 }\left( {\frac{1}{B_Z^2 }-\frac{1}{B_\theta ^2 }} 
\right)=0
\end{eqnarray}
Because of (\ref{eq30}): $f_1 +I_1 ={E_R } \mathord{\left/ {\vphantom {{E_R } 
{\left( {R\,B_Z } \right)}}} \right. \kern-\nulldelimiterspace} {\left( 
{R\,B_Z } \right)}$ equation (\ref{eq76}) yields an explicit algebraic expression 
for $E_R $ as a function of $s$ and the poloidal magnetic field components. This result is not compatible with $E_R $ given as an 
integral in the first equation of (\ref{eq30}), unless condition (\ref{eq32}) is satisfied. 
In this case follows from (\ref{eq74}): $f_1^2 =f_3^2 =\phi _0 '{\kern 1pt}^2$ and the potential becomes a function of $\psi $ only.

\vspace{1. cm}


\end{large}


\begin{thebibliography}{99}
\bibitem{kruskal}
{M. D. Kruskal, R. M. Kulsrud, The Physics of Fluids \textbf{1} (1958) 265}
\bibitem {newcomb} {W. A. Newcomb, The Physics of Fluids \textbf{2} (1959) 362}
\bibitem{pfirsch} {D. Pfirsch, A. Schl\"{u}ter, Max-Planck-Institut f\"{u}r Physik und Astrophysik, Report MPI/Pa/7/62 (1962) (unpublished)}
\bibitem{maschke} {E. K. Maschke, Plasma Physics \textbf{13} (1971) 905}
\bibitem{gates} {D. A. Gates, H. E. Mynick, R. B. White, Physics of Plasmas \textbf{11} (2004) L45}
\bibitem{grad} {R. L\"{u}st, A. Schl\"{u}ter, Zeitschrift f\"{u}r Naturforschung \textbf{12} (1957) 850 
\\ V. D. Shafranov, Sov. Phys. JETP \textbf{6} (1958) 545; Zh.Eksp.Teor. Fiz. \textbf{33} (1957) 710 
\\ H. Grad, H. Rubin, \textit{Proc. 2nd U. N. Int. Conf. on the Peaceful Uses of Atomic Energy} Geneva 1958, Vol. \textbf{31}, 190, Columbia University Press, New York (1959)}
\bibitem{callen}{X. Liu, J. D. Callen, C. G. Hegna, Physics of Plasmas \textbf{11} (2004) 4824L}
\bibitem{soloviev} {L. S. Soloviev, Sov. Phys. JETP \textbf{26} (1968) 400; Zh. Eksp. Teor. Fiz. \textbf{53} (1967) 626}
\bibitem{atanasiu} { C. V. Atanasiu, S. G\"{u}nter, K. Lackner, I. G. Miron, Physics of Plasmas \textbf {11} (2004) 3510}

\end{thebibliography}
\end{document}